
\def\cstok#1{\leavevmode\thinspace\hbox{\vrule\vtop{\vbox{\hrule\kern1pt
\hbox{\vphantom{\tt/}\thinspace{\tt#1}\thinspace}}
\kern1pt\hrule}\vrule}\thinspace}

\magnification=1200

\font\tit=cmbx10 scaled\magstep2

\font\abst=cmti9

\font\abstb=cmbx9

\font\addrt=cmti9

\font\aun=cmbx10


\font\ridbf=cmr9

\font\rauth=cmcsc10

\font\tas=cmti9


\topskip=1truecm
\abovedisplayskip=15truept plus 5truept minus 5truept
\belowdisplayskip=15truept plus 5truept minus 5truept
\newskip\myskip
\myskip=40truept
\clubpenalty=1000
\widowpenalty=1000
\hsize=6truein
\vsize=8.5truein
\newcount\who
\who=0
\rm
\def\centra#1{\vbox{\rightskip=0pt plus1fill\leftskip=0pt #1}}
\def\centrb#1{\vbox{\rightskip=0pt plus1fill\leftskip=\myskip #1}}
\def\titolo{}
\def\frac#1#2{{#1\over #2}}

\def\title#1{\vfill\eject
\ifodd\pageno\else\def\firstauthor{}\null\vfill\eject\fi
\firstp=\pageno
\def\speak{}
\parindent=0pt
\baselineskip=8truept
\line{\hfill\vbox{\tas\spaceskip=0pt plus1fil
\hbox to 62mm{Proceedings of the XI Italian Conference on}
\hbox to 62mm{General Relativity and Gravitational Physics}
\hbox to 62mm{Trieste (Italy), September 26--30, 1994}
\hbox to 62mm{\copyright 1995 World Scientific Publishing Company}}}

\baselineskip=20truept
\bigskip\medskip
\centra{\tit #1}
\bigskip\baselineskip=12pt\def\titolo{#1}}

\def\oneauthor#1#2{\bigskip\centrb{\aun #1}\medskip
\centrb{\addrt #2}\global\def\firstauthor{#1}}


\long\def\support#1#2{\footnote{}
{\noindent\line{\hbox to 6truecm{\hrulefill}\hfill}
\hbox to 20truept{\hfill$^{#1}$\ }{\abst #2}}}

\parskip=5pt
\long\def\summary#1{\bigskip
\vbox{\par\leftskip=\myskip\rightskip=0truept\baselineskip=11pt
\noindent{\abstb Summary:} \abst #1 \bigskip\centrb{\speak}}
\parindent=20truept
\write1{\string\record{\titolo}{\folio}{\firstauthor}}
}

\def\section#1#2{\who=1\bigskip\medskip\goodbreak{\bf\noindent\hbox to
20truept{#1.\hfil}#2}\nobreak
\medskip\nobreak\who=0}
\def\acknow{\bigskip\medskip\goodbreak{\bf Acknowledgements}\nobreak
\medskip\nobreak}
\def\subsection#1#2{\ifnum\who=0\bigskip\goodbreak\else\smallskip\fi
{
\noindent\hbox to 25truept{\ridbf #1.\hfil} \ridbf #2}\nobreak
\ifnum\who=0\medskip\fi\nobreak}
\def\references{\bigskip\medskip\goodbreak{\bf\noindent\hskip25truept
References}\nobreak
\medskip\nobreak
\frenchspacing\pretolerance=2000\parindent=25truept}

\def\paper#1#2#3#4#5#6{\item{\hbox to 20truept{[#1]\hfill}}
{\rauth #2,} {\it #3} {{\bf #4},} #5 (#6)\smallskip}
\def\book#1#2#3#4#5{\item{\hbox to 20truept{[#1]\hfill}} {\rauth #2,}
{\it #3}, #4 (#5)\smallskip}
\def\booknn#1#2#3#4#5{\item{\hbox to 20truept{\hfill}} {\rauth #2,}
{\it #3}, #4 (#5)\smallskip}
\def\papernn#1#2#3#4#5#6{\item{\hbox to 20truept{\hfill}}
{\rauth #2,} {\it #3} {{\bf #4},} #5 (#6)\smallskip}
\def\paperibidem#1#2#3#4#5#6#7#8#9{\item{\hbox to 20truept{[#1]\hfill}}
{\rauth #2,} {\it #3} {{\bf #4},} #5 (#6); ibid. {{\bf #7},} #8 (#9)
\smallskip}
\def\bookpre#1#2#3{\item{\hbox to 20truept{[#1]\hfill}} {\rauth #2,}
(#3)\smallskip}

\newcount\firstp

\headline={\ifnum\pageno=\firstp\line{\hfill}\else
\ifodd\pageno\vbox to 12pt{\line{\hfill{\it\titolo}\hfill\folio}%
\vfill\hrule}\else%
\vbox to 12pt{\line{\folio\hfill{\it \firstauthor}\hfill}\vfill\hrule}%
\fi\fi}

\footline={\ifnum\pageno=\firstp\line{\hfill\folio\hfill}\else%
\line{\hfill}\fi}

\title{Quantization of field theories in
the presence of boundaries}
\oneauthor{Giampiero Esposito}
{Istituto Nazionale di Fisica Nucleare, Sezione di
Napoli, Mostra d'Oltremare Padiglione 20, 80125 Napoli, Italy;
Dipartimento di Scienze Fisiche, Mostra d'Oltremare
Padiglione 19, 80125 Napoli, Italy.}

\summary{This paper reviews the progress made over the last
five years in studying boundary conditions and semiclassical
properties of quantum fields about 4-real-dimensional Riemannian
backgrounds. For massless spin-${1\over 2}$ fields one has a
choice of spectral or supersymmetric boundary conditions, and the
corresponding conformal anomalies have been evaluated by using
zeta-function regularization. For Euclidean Maxwell theory in
vacuum, the mode-by-mode analysis of BRST-covariant
Faddeev-Popov amplitudes has been performed for relativistic and
non-relativistic gauge conditions. For massless
spin-${3\over 2}$ fields, the contribution of physical degrees
of freedom to one-loop amplitudes, and the 2-spinor analysis
of Dirac and Rarita-Schwinger potentials, have been obtained.
In linearized gravity, gauge modes and ghost modes in the
de Donder gauge have been studied in detail. This program may
lead to a deeper understanding of different quantization
techniques for gauge fields and gravitation, to a new vision
of gauge invariance, and to new points of view in twistor
theory.}

\section{1}{Introduction}

The way in which quantum fields respond to the presence of
boundaries is responsible for many interesting physical
effects such as, for example, the Casimir effect, and the
quantization program of spinor fields, gauge fields and
gravitation in the presence of boundaries is currently leading
to a better understanding of modern quantum field theories.
The motivations for this investigation come from at least three
areas of physics and mathematics, i.e. [1-15]
\vskip 0.3cm
\noindent
(i) {\it Cosmology}. One wants to understand what is the quantum
state of the universe, and how to formulate boundary conditions
for the universe [16].
\vskip 0.3cm
\noindent
(ii) {\it Field Theory}. It appears necessary to get a deeper
understanding of different quantization techniques in field
theory, i.e. the reduction to physical degrees of freedom
before quantization, or the Faddeev-Popov Lagrangian method,
or the Batalin-Fradkin-Vilkovisky extended phase space.
Moreover, perturbative properties of supergravity theories
and conformal anomalies in field theory deserve further thinking,
especially within the framework of semiclassical evaluation
of path integrals in field theory via zeta-function
regularization.
\vskip 0.3cm
\noindent
(iii) {\it Mathematics}. A (pure) mathematician may regard
quantum cosmology as a problem in cobordism theory, and
one-loop quantum cosmology as a relevant application of
the theory of eigenvalues in Riemannian geometry, of
self-adjointness theory, and of the analysis of asymptotic
heat kernels for manifolds with boundary.

On using zeta-function regularization [1-15], the $\zeta(0)$
value yields the scaling of quantum amplitudes and the one-loop
divergences of physical theories. The choices to be made concern
the quantization technique, the background 4-geometry, the
boundary 3-geometry, the boundary conditions respecting
Becchi-Rouet-Stora-Tyutin invariance and local supersymmetry,
the gauge condition, the regularization algorithm [9]. By the
latter, we mean that $\zeta(0)$ can be evaluated by means of
analytic and direct techniques, i.e. Laplace transform of the
heat kernel and Euler-Maclaurin formula, or Laplace and
Watson transforms, or zeta-function at large $x$, or BKKM
formalism [1-15,17]. However, $\zeta(0)$ can also be obtained
in a geometric (though indirect) way, by evaluating the
corresponding coefficient in the Schwinger-DeWitt asymptotic
expansion.

The following sections describe recent progress on these issues,
starting from massless spin-${1\over 2}$ fields, and then
focusing on Euclidean Maxwell theory, spin-${3\over 2}$
potentials and linearized gravity. Open problems are presented
in section 6.

\section{2}{Massless spin-${1\over 2}$ fields}

The early analysis in [1-3] studied a massless spin-${1\over 2}$
field at one-loop about flat Euclidean 4-space bounded by a
3-sphere of radius $a$.
By virtue of the first-order nature of the Dirac
operator, an elliptic operator mapping elements of primed
spin-space to unprimed spin-space (and the other way around)
in even dimension, one has a choice of {\it spectral}
or {\it local} boundary conditions. Using two-component spinor
notation as in [1-3], the massless spin-${1\over 2}$ field is
represented by a pair of independent spinor fields
$\Bigr(\psi^{A},{\widetilde \psi}^{A'}\Bigr)$. Their expansion on
a family of 3-spheres centred on the origin takes the form
[1-3]
$$
\psi^{A}=
{\tau^{-{3\over 2}}\over 2\pi}
\sum_{n=0}^{\infty}\sum_{p,q=1}^{(n+1)(n+2)}
\alpha_{n}^{pq}\Bigr[m_{np}(\tau)\rho^{nqA}
+{\widetilde r}_{np}(\tau){\overline \sigma}^{nqA}\Bigr]
\; \; \; \; ,
\eqno (2.1)
$$
$$
{\widetilde \psi}^{A'}=
{\tau^{-{3\over 2}}\over 2\pi}
\sum_{n=0}^{\infty}\sum_{p,q=1}^{(n+1)(n+2)}
\alpha_{n}^{pq}\left[{\widetilde m}_{np}(\tau)
{\overline \rho}^{nqA'}
+r_{np}(\tau)\sigma^{nqA'}\right]
\; \; \; \; ,
\eqno (2.2)
$$
where $\alpha_{n}^{pq}$ are block-diagonal matrices with
blocks $\pmatrix {1&1\cr 1&-1 \cr}$, and the $\rho$- and
$\sigma$-harmonics obey the identities described in [1-3].
In the classical theory, since the field is massless, the
modes $m_{np}(\tau)$ and $r_{np}(\tau)$ are regular
$\forall \tau \in [0,a]$, while the modes
${\widetilde m}_{np}(\tau)$ and ${\widetilde r}_{np}(\tau)$
behave as negative powers of the Euclidean-time coordinate,
and hence are singular at the origin of Euclidean 4-space.
Thus, to find a smooth solution of the classical
boundary-value problem, only half of the fermionic field can
be fixed at the boundary, corresponding to the modes
$m_{np}$ and $r_{np}$ multiplying harmonics having positive
eigenvalues for the intrinsic 3-dimensional Dirac operator
at the boundary [1-3]. This part of the spin-${1\over 2}$
field is denoted with the label $(+)$, while the singular
part is denoted by the label $(-)$. The expansions (2.1)-(2.2)
are then re-expressed as
$$
\psi^{A}=\psi_{(+)}^{A}+\psi_{(-)}^{A}
\; \; \; \; ,
\eqno (2.3)
$$
$$
{\widetilde \psi}^{A'}={\widetilde \psi}_{(-)}^{A'}
+{\widetilde \psi}_{(+)}^{A'}
\; \; \; \; .
\eqno (2.4)
$$
In the case of spectral boundary conditions, we impose [1,3]
$$
\Bigr[\psi_{(+)}^{A}\Bigr]_{\partial M}=0
\; \; \; \; , \; \; \; \;
\Bigr[{\widetilde \psi}_{(+)}^{A'}\Bigr]_{\partial M}=0
\; \; \; \; .
\eqno (2.5)
$$

In the case of local boundary conditions, motivated by local
supersymmetry and supergravity multiplets [1-2], we
require instead that the spinor field defined as
(here $\epsilon \equiv \pm 1$)
$$
\Phi^{A'} \equiv \sqrt{2} \; {_{e}n_{A}^{\; \; A'}}
\; \psi^{A}- \epsilon \; {\widetilde \psi}^{A'}
\; \; \; \; ,
\eqno (2.6)
$$
should vanish at the boundary, i.e.
$$
\Bigr[\Phi^{A'}\Bigr]_{\partial M}=0
\; \; \; \; .
\eqno (2.7)
$$
In [1-2] the existence was proved of self-adjoint extensions
of the Dirac operator subject to local boundary conditions,
after generalizing a result for complex scalar fields due to
von Neumann. Remarkably, for massless spin-${1\over 2}$ fields,
one finds $\zeta(0)={11\over 360}$ in the case of flat Euclidean
4-space bounded by a 3-sphere, both for spectral and local
boundary conditions, although the spectra are different. This
result was first proved in [1-3], and then confirmed with the
help of the more powerful technique used in [4-5]. It seems to
reflect a symmetry of the classical boundary-value problem, as
shown in [18]. In [18] we have also worked out the boundary term
necessary in the action functional when the spinor field defined
in (2.6) is not set to zero at the boundary. The corresponding
Euclidean action is [18]
$$ \eqalignno{
I_{E}&={i\over 2}\int_{M}\left[{\widetilde \psi}^{A'}
\Bigr(\nabla_{AA'} \; \psi^{A}\Bigr)
-\Bigr(\nabla_{AA'} \; {\widetilde \psi}^{A'}\Bigr)
\psi^{A}\right]\sqrt{{\rm det} \; g} \; d^{4}x \cr
&+{i \epsilon \over 2}\int_{\partial M}
\Phi^{A'} \; {_{e}n_{AA'}} \; \psi^{A}
\sqrt{{\rm det} \; h} \; d^{3}x
\; \; \; \; .
&(2.8)\cr}
$$

\section{3}{Euclidean Maxwell theory}

We are interested in the mode-by-mode analysis of
BRST-covariant Faddeev-Popov amplitudes, which relies
on the expansion of the electromagnetic potential in
harmonics on the boundary 3-geometry. In the case of
3-sphere boundaries, one has [1,6-9]
$$
A_{0}(x,\tau)=\sum_{n=1}^{\infty}R_{n}(\tau)Q^{(n)}(x)
\; \; \; \; ,
\eqno (3.1)
$$
$$
A_{k}(x,\tau)=\sum_{n=2}^{\infty}
\biggr[f_{n}(\tau)S_{k}^{(n)}(x)
+g_{n}(\tau)P_{k}^{(n)}(x)\biggr]
\; \; \; \; ,
\eqno (3.2)
$$
where $Q^{(n)}(x),S_{k}^{(n)}(x)$ and $P_{k}^{(n)}(x)$
are scalar, transverse and longitudinal vector harmonics
on $S^{3}$ respectively.

Magnetic boundary conditions set to zero at the boundary the
gauge-averaging function, the tangential components of the
potential, and the ghost field, i.e.
$$
\Bigr[\Phi(A)\Bigr]_{\partial M}=0 \; \; \; \; ,
\; \; \; \; \Bigr[A_{k}\Bigr]_{\partial M}=0
\; \; \; \; , \; \; \; \;
[\epsilon]_{\partial M}=0
\; \; \; \; .
\eqno (3.3)
$$
Moreover, electric conditions set to zero at the boundary the
normal component of the potential,
the partial derivative with respect
to $\tau$ of the tangential components of the potential,
and the normal derivative of the ghost field, i.e.
$$
\Bigr[A_{0}\Bigr]_{\partial M}=0
\; \; \; \; , \; \; \; \;
\Bigr[\partial A_{k} / \partial \tau\Bigr]_{\partial M}=0
\; \; \; \; , \; \; \; \;
\Bigr[\partial \epsilon / \partial n\Bigr]_{\partial M}=0
\; \; \; \; .
\eqno (3.4)
$$
One may check that these boundary conditions are compatible
with BRST transformations, and do not give rise to additional
boundary conditions after a gauge transformation.

By using zeta-function regularization and flat Euclidean
backgrounds, the effects of relativistic gauges are as
follows [1,6-8].
\vskip 0.3cm
\noindent
(i) In the Lorentz gauge, the mode-by-mode analysis of one-loop
amplitudes agrees with the results of the Schwinger-DeWitt
technique, both in the 1-boundary case (i.e. the disk) and
in the 2-boundary case (i.e. the ring).
\vskip 0.3cm
\noindent
(ii) In the presence of boundaries, the effects of gauge modes
and ghost modes {\it do not} cancel each other.
\vskip 0.3cm
\noindent
(iii) When combined with the contribution of physical degrees
of freedom, i.e. the transverse part of the potential, this lack
of cancellation is exactly what one needs to achieve agreement
with the results of the Schwinger-DeWitt technique.
\vskip 0.3cm
\noindent
(iv) Thus, physical degrees of freedom are, by themselves,
insufficient to recover the full information about one-loop
amplitudes.
\vskip 0.3cm
\noindent
(v) Moreover, even on taking into account physical, non-physical
and ghost modes, the analysis of relativistic gauges different
from the Lorentz gauge yields gauge-invariant amplitudes
only in the 2-boundary case.
\vskip 0.3cm
\noindent
(vi) Gauge modes obey a coupled set of second-order eigenvalue
equations (section 6). For some particular choices of gauge
conditions it is possible to decouple such a set of differential
equations, by means of two functional matrices which diagonalize
the original operator matrix.
\vskip 0.3cm
\noindent
(vii) For arbitrary choices of relativistic gauges, gauge modes
remain coupled. The explicit proof of gauge invariance of
quantum amplitudes becomes a problem in homotopy theory.
Hence there seems to be a deep relation between the
Atiyah-Patodi-Singer theory of Riemannian 4-manifolds with
boundary [19], the zeta-function, and the BKKM function [17]:
$$
I(M^{2},s) \equiv \sum_{n=n_{0}}^{\infty}
d(n) \; n^{-2s} \; \log \Bigr[f_{n}(M^{2})\Bigr]
\; \; \; \; .
\eqno (3.5)
$$

In (3.5), $d(n)$ is the degeneracy of the eigenvalues
parametrized by the integer $n$, and $f_{n}(M^{2})$ is the
function occurring in the equation obeyed by the eigenvalues
by virtue of boundary conditions, after taking out fake roots.
The analytic continuation of (3.5) to the whole complex-$s$ plane
is given by [17]
$$
``I(M^{2},s)"={I_{\rm pole}(M^{2})\over s}
+I^{R}(M^{2})+O(s)
\; \; \; \; ,
\eqno (3.6)
$$
and enables one to evaluate $\zeta(0)$ as [17]
$$
\zeta(0)=I_{\rm log}+I_{\rm pole}(\infty)-I_{\rm pole}(0)
\; \; \; \; ,
\eqno (3.7)
$$
$I_{\rm log}$ being the coefficient of $\log(M)$ appearing in
$I^{R}$ as $M \rightarrow \infty$.

\section{4}{Spin-${3\over 2}$ potentials}

In the early analysis in [1-5,10], it was found that the
contribution of physical degrees of freedom to the full
$\zeta(0)$ for gravitinos is equal to $-{289\over 360}$
both for spectral and local boundary conditions, in the
case of flat Euclidean 4-space bounded by a 3-sphere.
More recently, attention has been focused on some basic
properties of Dirac and Rarita-Schwinger potentials for
spin ${3\over 2}$, motivated by the attempt of Roger Penrose
to define twistors as spin-${3\over 2}$ charges [11-14].
Following [11-14], Tables I and II
present primary and secondary
potentials in the Dirac and Rarita-Schwinger schemes, with
their gauge transformations. In [11-12] the two-component
spinor analysis of the four potentials of the totally
symmetric and independent field strengths for spin ${3\over 2}$
has been applied to the case of a 3-sphere boundary. It has been
shown that the Breitenlohner-Freedman-Hawking reflective
boundary conditions:
$$
2^{s} \; {_{e}n^{AA'}} ... {_{e}n^{LL'}} \;
\phi_{A...L}=\pm
{\widetilde \phi}^{A'...L'}
\; \; \; \; {\rm at} \; \partial M
\eqno (4.1)
$$
can only be imposed in a flat Euclidean background, for which
the gauge freedom in the choice of the potentials remains.

More recently, the Rarita-Schwinger field equations have been
studied in four-real-dimensional Riemannian backgrounds with
boundary [13-14], subject to the Luckock-Moss-Poletti boundary
conditions compatible with local supersymmetry:
$$
\sqrt{2} \; {_{e}n_{A}^{\; \; A'}} \; \psi_{\; \; i}^{A}
=\pm {\widetilde \psi}_{\; \; i}^{A'}
\; \; \; \; {\rm at} \; \partial M
\; \; \; \; .
\eqno (4.2)
$$
Gauge transformations on the potentials are compatible with
the field equations providing the background is Ricci-flat,
in agreement with previous results in the literature.
However, the preservation of boundary conditions under such
gauge transformations leads to a restriction of the gauge
freedom. The recent construction by Penrose of secondary
potentials which supplement the Rarita-Schwinger potentials,
jointly with the boundary conditions (4.2),
has been shown to imply that the background 4-geometry is
further restricted to be totally flat [13].

\section{5}{Linearized gravity}

A detailed mode-by-mode study of perturbative quantum gravity
about a flat Euclidean background bounded by two concentric
3-spheres, including non-physical degrees of freedom and ghost
modes, leads to one-loop amplitudes in agreement with the
covariant Schwinger-DeWitt method [15]. This calculation provides
the generalization of the previous analysis of fermionic fields
and electromagnetic fields [1-9]. The basic idea is to expand
the metric perturbations $h_{00},h_{0i}$ and $h_{ij}$ on a
family of 3-spheres centred on the origin, and then use the
de Donder gauge-averaging function in the Faddeev-Popov
Euclidean action. The resulting eigenvalue equation for metric
perturbations about a flat Euclidean background:
$$
\cstok{\ } h_{\mu \nu}+\lambda \; h_{\mu \nu}=0
\; \; \; \; ,
\eqno (5.1)
$$
gives rise to seven coupled eigenvalue equations for metric
perturbations. On considering also the ghost 1-form
$\varphi_{\mu}$, and imposing mixed boundary conditions on
metric and ghost perturbations [15]
$$
\Bigr[h_{ij}\Bigr]_{\partial M}
= \Bigr[h_{0i}\Bigr]_{\partial M}
=\Bigr[\varphi_{0}\Bigr]_{\partial M}=0
\; \; \; \; ,
\eqno (5.2)
$$
$$
\left[{\partial h_{00}\over \partial \tau}
+{6\over \tau}h_{00}-{\partial \over \partial \tau}
\Bigr(g^{ij}h_{ij}\Bigr)\right]_{\partial M}=0
\; \; \; \; ,
\eqno (5.3)
$$
$$
\left[{\partial \varphi_{i}\over \partial \tau}
-{2\over \tau} \varphi_{i}\right]_{\partial M}=0
\; \; \; \; ,
\eqno (5.4)
$$
the analysis in [15] has shown that the full $\zeta(0)$
vanishes in the 2-boundary problem, while the contributions
of ghost modes and gauge modes {\it do not} cancel each
other, as it already happens for Euclidean Maxwell theory
(section 3).

\section{6}{Open problems}

The main open problem seems to be the explicit proof of gauge
invariance of one-loop amplitudes for relativistic gauges, in
the case of flat Euclidean space bounded by two concentric
3-spheres. For this purpose, one may have to show that, for
coupled gauge modes, $I_{\rm log}$ and the difference
$I_{\rm pole}(\infty)-I_{\rm pole}(0)$ are not affected by a
change in the gauge parameters. Three steps are in order:
\vskip 0.3cm
\noindent
(i) To relate the regularization at large $x$
used in [1] to the BKKM regularization.
\vskip 0.3cm
\noindent
(ii) To evaluate $I_{\rm log}$ from an asymptotic analysis
of coupled eigenvalue equations.
\vskip 0.3cm
\noindent
(iii) To evaluate $I_{\rm pole}(\infty)-I_{\rm pole}(0)$ by
relating the analytic continuation to the whole complex-$s$
plane of the difference $I(\infty,s)-I(0,s)$,
to the analytic continuation of the zeta-function.

The last step may involve a non-local transform
relating the BKKM function to the zeta-function, and a
non-trivial application of the Atiyah-Patodi-Singer theory
of Riemannian 4-manifolds with boundary [19]. In other words,
one might have to prove that, {\it in the 2-boundary problem
only}, $I_{\rm pole}(\infty)-I_{\rm pole}(0)$ resulting
from coupled gauge modes is the residue of a meromorphic
function, invariant under a smooth variation in the gauge
parameters of the matrix of elliptic self-adjoint operators
appearing in the system [8]
$$
{\widehat {\cal A}}_{n}g_{n}+{\widehat {\cal B}}_{n}R_{n}=0
\; \; \; \; , \; \; \; \;
\forall n \geq 2
\; \; \; \; ,
\eqno (6.1)
$$
$$
{\widehat {\cal C}}_{n}g_{n}+{\widehat {\cal D}}_{n}R_{n}=0
\; \; \; \; , \; \; \; \;
\forall n \geq 2
\; \; \; \; .
\eqno (6.2)
$$
Here, denoting by $\gamma_{1},\gamma_{2},\gamma_{3}$ three
dimensionless parameters which enable one to write the most
general gauge-averaging function, and by $\alpha$ the positive
dimensionless parameter occurring in the Faddeev-Popov Euclidean
action, one has [8]
$$
{\widehat {\cal A}}_{n} \equiv
{d^{2}\over d\tau^{2}}+{1\over \tau}{d\over d\tau}
-{\gamma_{3}^{2}\over \alpha}{(n^{2}-1)\over \tau^{2}}
+\lambda_{n}
\; \; \; \; ,
\eqno (6.3)
$$
$$
{\widehat {\cal B}}_{n} \equiv
- \Bigr(1+{\gamma_{1}\gamma_{3}\over \alpha}\Bigr)
(n^{2}-1){d\over d\tau}
- \Bigr(1+{\gamma_{2}\gamma_{3}\over \alpha}\Bigr)
{(n^{2}-1)\over \tau}
\; \; \; \; ,
\eqno (6.4)
$$
$$
{\widehat {\cal C}}_{n} \equiv
\Bigr(1+{\gamma_{1}\gamma_{3}\over \alpha}\Bigr)
{1 \over \tau^{2}}{d\over d\tau}
+{\gamma_{3}\over \alpha}(\gamma_{1}-\gamma_{2})
{1\over \tau^{3}}
\; \; \; \; ,
\eqno (6.5)
$$
$$
{\widehat {\cal D}}_{n} \equiv
{\gamma_{1}^{2}\over \alpha}{d^{2}\over d\tau^{2}}
+{3\gamma_{1}^{2}\over \alpha}{1\over \tau}
{d\over d\tau}
+\left[{\gamma_{2}\over \alpha}(2\gamma_{1}-\gamma_{2})
-(n^{2}-1)\right]{1\over \tau^{2}}+\lambda_{n}
\; \; \; \; .
\eqno (6.6)
$$

Other relevant research problems are the mode-by-mode analysis
of one-loop amplitudes for gravitinos, including gauge modes
and ghost modes studied within the Faddeev-Popov formalism,
and the study of gauge transformations for the secondary potentials
$\rho_{A'}^{\; \; \; CB}$ and $\theta_{A}^{\; \; C'B'}$ which
supplement Rarita-Schwinger potentials (section 4). In that model
it is not yet clear whether there is an underlying global theory,
what parts of the curvature are the obstructions to defining
a global theory, what are the key features of the global theory
(if it exists). Moreover, it appears necessary to understand
whether one can define twistors as charges for spin ${3\over 2}$
in a Riemannian background which is not Ricci-flat, and whether
one can reconstruct the Riemannian 4-world from the resulting
twistor space, or from whatever mathematical structure is going
to replace twistor space.

Last, but not least, the mode-by-mode analysis of linearized gravity
in the de Donder gauge in the 1-boundary case, the unitary gauge
for linearized gravity, and the mode-by-mode analysis of
one-loop amplitudes in the case of curved backgrounds, appear
to be necessary to complete the picture outlined so far.
The recent progress on problems with boundaries,
however, seems to strengthen the evidence
in favour of new perspectives being in sight in quantum field
theory and quantum cosmology.

\acknow

I am indebted to P. D. D'Eath, A. Yu. Kamenshchik, I. V. Mishakov
and G. Pollifrone for collaboration on many topics described in
this review paper.

\vskip 1truecm

$${\vbox{
\offinterlineskip\halign{\strut\vrule#&
\hfil\quad#\quad\hfil&\vrule#&
\hfil\quad#\quad\hfil&\vrule#&
\hfil\quad#\quad\hfil&\vrule#\cr
\multispan{7}\hfil TABLE I\hfil\cr
\noalign{\bigskip}
\multispan{7}\vrule\hrulefill\vrule\cr
&&&&&&\cr
&{$\;$}&&{Dirac Potentials}&&{Gauge Freedom}&\cr
&&&&&&\cr
\multispan{7}\vrule\hrulefill\vrule\cr
&&&&&&\cr
&Primary&&$\phi_{ABC}=\nabla_{CC'} \; \Gamma_{\; \; \; AB}^{C'}$&&
${\widehat \Gamma}_{\; \; \; AB}^{C'} \equiv
\Gamma_{\; \; AB}^{C'}+\nabla_{A}^{\; \; C'} \; \nu_{B}$&\cr
&&&&&&\cr
\multispan{7}\vrule\hrulefill\vrule\cr
&&&&&&\cr
&Primary&&${\widetilde \phi}_{A'B'C'}=\nabla_{CC'} \;
\gamma_{\; \; A'B'}^{C}$&&
${\widehat \gamma}_{\; \; A'B'}^{C} \equiv
\gamma_{\; \; A'B'}^{C}+\nabla_{\; \; A'}^{C} \;
{\widetilde \nu}_{B'}$&\cr
&&&&&&\cr
\multispan{7}\vrule\hrulefill\vrule\cr
&&&&&&\cr
&Secondary&&$\Gamma_{\; \; \; AB}^{C'}=\nabla_{BB'} \;
\Lambda_{A}^{\; \; B'C'}$&&
${\widehat \Lambda}_{A}^{\; \; B'C'} \equiv
\Lambda_{A}^{\; \; B'C'}+\nabla_{\; \; A}^{B'} \;
{\widetilde \chi}^{C'}$&\cr
&&&&&&\cr
\multispan{7}\vrule\hrulefill\vrule\cr
&&&&&&\cr
&Secondary&&$\gamma_{\; \; A'B'}^{C}=\nabla_{BB'} \;
\rho_{A'}^{\; \; \; BC}$&&
${\widehat \rho}_{A'}^{\; \; \; BC} \equiv
\rho_{A'}^{\; \; \; BC}+\nabla_{\; \; A'}^{B}
\; \chi^{C}$&\cr
&&&&&&\cr
\multispan{7}\vrule\hrulefill\vrule\cr
}}}$$

\vskip 1truecm

$${\vbox{\offinterlineskip\halign{\strut\vrule#&
\hfil\quad#\quad\hfil&\vrule#&
\hfil\quad#\quad\hfil&\vrule#&
\hfil\quad#\quad\hfil&\vrule#\cr
\multispan{7}\hfill TABLE II\hfill\cr
\noalign{\bigskip}
\multispan{7}\vrule\hrulefill\vrule\cr
&&&&&&\cr
&{$\;$}&&{Rarita-Schwinger Potentials}&&{Gauge Freedom}&\cr
&&&&&&\cr
\multispan{7}\vrule\hrulefill\vrule\cr
&&&&&&\cr
&Primary&&$\psi_{A \; \mu}=\Gamma_{\; \; \; AB}^{C'}
\; e_{\; \; C'\mu}^{B}$&&
${\widehat \Gamma}_{\; \; \; BC}^{A'} \equiv
\Gamma_{\; \; \; BC}^{A'}+\nabla_{\; \; B}^{A'}
\; \nu_{C}$&\cr
&&&&&&\cr
\multispan{7}\vrule\hrulefill\vrule\cr
&&&&&&\cr
&Primary&&${\widetilde \psi}_{A' \; \mu} = \gamma_{\; \; A'B'}^{C}
\; e_{C \; \; \; \; \mu}^{\; \; \; B'}$&&
${\widehat \gamma}_{\; \; B'C'}^{A} \equiv
\gamma_{\; \; B'C'}^{A}+\nabla_{\; \; B'}^{A} \; \lambda_{C'}$&\cr
&&&&&&\cr
\multispan{7}\vrule\hrulefill\vrule\cr
&&&&&&\cr
&Secondary&&$\Gamma_{AB}^{\; \; \; \; \; C'}
=\nabla_{B'B} \; \theta_{A}^{\; \; C'B'}$&&
${\widehat \theta}_{A}^{\; \; A'B'} \equiv
\theta_{A}^{\; \; A'B'}+\nabla_{A}^{\; \; A'} \; \mu^{B'}$&\cr
&&&&&&\cr
\multispan{7}\vrule\hrulefill\vrule\cr
&&&&&&\cr
&Secondary&&$\gamma_{A'B'}^{\; \; \; \; \; \; \; C}=\nabla_{BB'} \;
\rho_{A'}^      {\; \; \; CB}$&&
${\widehat \rho}_{A'}^{\; \; \; AB} \equiv
\rho_{A'}^{\; \; \; AB}+\nabla_{A'}^{\; \; \; A} \; \xi^{B}$&\cr
&&&&&&\cr
\multispan{7}\vrule\hrulefill\vrule\cr
}}}$$

\references

\book{1}{G. Esposito}{Quantum Gravity, Quantum Cosmology and
Lorentzian Geometries}{Lecture Notes in Physics, New Series
m: Monographs, Vol. m 12, second corrected and enlarged
edition (Berlin: Springer-Verlag)}{1994}
\paper{2}{P. D. D'Eath and G. Esposito}{Phys. Rev.}
{D 43}{3234}{1991}
\paper{3}{P. D. D'Eath and G. Esposito G.}{Phys. Rev.}
{D 44}{1713}{1991}
\paper{4}{A. Yu. Kamenshchik and I. V. Mishakov}{Phys. Rev.}
{D 47}{1380}{1993}
\paper{5}{A. Yu
Kamenshchik and I. V. Mishakov}{Phys. Rev.}
{D 49}{816}{1994}
\paper{6}{G.
Esposito}{Class. Quantum Grav.}{11}{905}{1994}
\paper{7}{G. Esposito, A. Yu. Kamenshchik, I. V. Mishakov and G.
Pollifrone}{Class. Quantum Grav.}{11}{2939}{1994}
\book{8}{G. Esposito, A. Yu. Kamenshchik, I. V. Mishakov and G.
Pollifrone}{Relativistic Gauge Conditions in Quantum
Cosmology}{}{DSF preprint 95/8}
\paper{9}{G. Esposito and A. Yu. Kamenshchik}{Phys. Lett.}
{B 336}{324}{1994}
\paper{10}{G. Esposito}{Int. J. Mod. Phys.} {D 3}{593}{1994}
\paper{11}{G. Esposito and G. Pollifrone}{Class. Quantum
Grav.}{11}{897}{1994}
\book{12}{G. Esposito and G. Pollifrone}{Twistors and
Spin-${3\over 2}$ Potentials in Quantum Gravity}{
in {\it Twistor Theory}, ed. S. Huggett (New York:
Marcel Dekker) p 35}{1994}
\book{13}{G. Esposito, G. Gionti, A. Yu. Kamenshchik, I. V. Mishakov
and G. Pollifrone}{Spin-${3\over 2}$ Potentials in
Backgrounds with Boundary}{}{DSF preprint 95/20}
\book{14}{G. Esposito}{Complex General Relativity}
{Fundamental Theories of Physics, Vol. 69 (Dordrecht: Kluwer
Academic)}{1995}
\paper{15}{G. Esposito, A. Yu. Kamenshchik, I. V. Mishakov and G.
Pollifrone}{Phys. Rev.}{D 50}{6329}{1994}
\paper{16}{S. W. Hawking}{Nucl. Phys.}{B 239}{257}{1994}
\paper{17}{A. O. Barvinsky, A. Yu. Kamenshchik and I. P. Karmazin}
{Ann. Phys.}{219}{201}{1992}
\paper{18}{G. Esposito, H. A. Morales-T\'ecotl and G. Pollifrone}
{Found. Phys. Lett.}{7}{303}{1994}
\paper{19}{M. F. Atiyah, V. K. Patodi and I. M. Singer}
{Math. Proc. Camb. Phil. Soc.}{79}{71}{1976}

\bye